\documentclass[a4paper,aps,twoside,prd,amsmath,nofootinbib,
superscriptaddress,showpacs,showkeys]{revtex4-1}
\pdfoutput=1
\usepackage{amsmath}
\usepackage{natbib}
\usepackage{graphicx}
\usepackage{enumitem}
\usepackage{xcolor}
\newcommand{\Mov}[1]{{\color{black}{#1}}}

\begin{document}
\title{Surface Density of Disk Galaxies in MOND}  



\author{Antonino Del Popolo 
}



\address{%
  Dipartimento di Fisica e Astronomia, University Of Catania, Viale Andrea Doria 6, 95125 Catania, Italy; antonino.delpopolo@unict.it}
\address{
  Institute of Astronomy, Russian Academy of Sciences, Pyatnitskaya Str., 48, 119017 Moscow, Russia}

 \author{ Morgan Le Delliou 
} 
%
\email{Correspondence: morgan.ledelliou.ift@gmail.com}
\address{
  Institute of Theoretical Physics, School of Physical Science and Technology, Lanzhou University, No. 222, South Tianshui Road, Lanzhou 730000, China}
\address{
  Lanzhou Center for Theoretical Physics, Key Laboratory of Theoretical Physics of Gansu Province, Lanzhou University, Lanzhou 730000, China}
\address{
  Instituto de Astrof\'isica e Ci\^encias do Espa\c co, Universidade de Lisboa, Faculdade de Ci\^encias, Ed. C8, Campo Grande, 1769-016 Lisboa, Portugal}

\date{
}


\begin{abstract}In this paper, we extend a paper by Milgrom (2009, MNRAS 398, 1023) dealing with the existence of a quasi-universal surface density for object of all mass and structure, if they are in the Newtonian regime, i.e., that their mean acceleration is larger than MOND typical acceleration $a_0$. This result is in agreement with Donato et al. (2009)'s results, claiming the existence of a quasi-universal surface density in all masses in galaxies. 
The Milgrom paper also predicts that objects with mean inner acceleration smaller than the values discussed do 
not show the quasi-universal behavior of the surface density discussed.  In the present paper, we extend the result of Milgrom's paper, based on a point mass model, considering spiral galaxies, modelled with a double exponential disk. Similar to Milgrom's results, we find the existence of a universal surface density for galaxies with large surface density, and a different behavior for galaxies having small surface density. 
\end{abstract}

\keywords{galaxies; alternative theory of gravity; galaxies surface density} 
%
%
%
\maketitle

\section{Introduction}

The $\Lambda$CDM model is able to predict with high accuracy the observations on cosmological scales\footnote{we recall that at cosmological scales the $\Lambda$CDM paradigm is affected by the cosmological constant problem \citep{Weinberg,Astashenok}.}, and~intermediate scales \cite{Spergel,Kowalski,Percival,Komatsu,2013AIPC15482D,2014IJMPD2330005D}. However, it has problems in describing the observations on small scales from tens of parsecs to some kiloparsecs.
These problems are coined the 
so-called "small scale problems of the $\Lambda$CDM" model. One of these problems, dubbed the ``Cusp/Core'' problem, is the discrepancy between 
cuspy density profiles of galaxies in N-body simulations \citep{nfw1996,nfw1997,Navarro2010}, and~the observed cored profiles in dwarf spirals, dwarf spheroidals (dSphs), and~Low Surface Brightness (LSB) galaxies
\citep{Moore1994,Flores1994,Burkert1995,Swatersetal2003,DelPopolo2009,2012MNRAS419971D,deBloketal2003,DelPopoloKroupa2009,DelPopoloHiotelis2014}
. Another problem is the ``missing satellite problem'' dealing with the discrepancy between the number of subhalos predicted in N-body simulations \citep{Klypin1999,Moore1999}, and~the one observed. A~third 
problem is the ``too-big-to-fail'' problem, i.e.,~the subhaloes are too dense 
compared to what we observe around the Milky Way \citep{GarrisonKimmel2013,GarrisonKimmel2014}. 

This flurry of issues remaining with the $\Lambda$CDM model has led to a debate on the validity of the dark matter (hereafter DM) assumption, as~found in a Special Issue of Universe~\cite{galaxies8010009,galaxies8010012,galaxies8010019,galaxies8020035,galaxies8020036,galaxies8020037,galaxies8020042,galaxies8020047,galaxies8020048,galaxies8030054}. The~debate remains~vivid.
 
To solve the 
problems mentioned above, it 
has been proposed to 
modify the particles constituting DM \citep{2000ApJ542622C,2001ApJ551608S,2000NewA5103G,2000ApJ534L127P}, to~modify the power spectrum  (e.g., \citep{2003ApJ59849Z}), to~modify the theory of gravity (\citep{1970MNRAS1501B,1980PhLB9199S,1983ApJ270365M,1983ApJ270371M,Ferraro2012}, including MOND, the~theory used in this work), or~astrophysical solutions based on mechanisms that ``heat'' DM (a) supernovae feedback, and~(b) transfer of energy and angular momentum from baryons to DM 
through dynamical~friction.

In this context, using as a fitting profile a pseudo-isothermal profile, obtained through mass-modeling of the rotation curves of 55 galaxies~\cite{Kormendy2004}, several interesting relations among the DM halos~parameters were found. 

Among the quantities by them introduced, 
$\mu_{0D} =\rho_0 r_0$, 
is a sort of surface density, where $r_0$ is the core radius of the pseudo-isothermal profile, and~$\rho_0$ its central density. In~the case of late galaxies, they found that $\mu_{0D}$ was independent from galaxy luminosity, and~they found a value of $\simeq 100 M_{\odot}/pc^{2}$. The~previous result was extended to $\simeq  1000$ galaxies (spirals, dwarfs, ellipticals, etc.) by (\cite{Donato}, hereafter D09)
.
This new work 
found again a quasi-universality of the central surface density of DM~halos.  

As already reported, Milgrom~\cite{Milgrom2009} used MOND, a~modified gravity set at the level of Newtonian dynamics, to~show that in the Newtonian regime $\mu_{0D}$ has a quasi-universal value (\cite{Gentile2009}, hereafter G09). 
showed that the quasi-universality was present in the surface density of the luminous~matter. 

One issue to stress is that 
D09 and G09 assume that all galaxies (from dwarfs to ellipticals) have a flat inner spherical profile (note that $r$ in this work designates the spherical radius) described by a Burkert DM halo density profile:
\begin{equation}
\rho(r)=\frac{\rho_0 r_0^3}{(r+r_0)(r^2+r_0^2)}.
\end{equation}

In reality, it is well known that the Burkert profile usually gives a good fit to the rotation curves of dwarfs and LSBs (\citep{Gentile2004,Gentile2007,DelPopolo2009}, but~this is not true for giant galaxies, and~ellipticals, and~moreover there are 
several exceptions \citep{Simon2005,THINGS,2012MNRAS42438D,Simon2005}). 

Since not all galaxies are fitted by a Burkert profile and have 
an inner flat profile, 
the previous discussion raises obvious doubts as to the D09 and~G09 conclusions. In~fact, several authors raised doubts as to the D09 and~G09 results, concluding that
the surface density is not~universal. 

For example, Napolitano et al. \cite{Napolitano2010} showed that in the case of local early type galaxies
the projected density within the effective radius is larger than that of dwarfs and~spirals.

Ref.~\cite{Boyarsky} used a much larger sample of that of D09, and~G09, and~showed a systematic increase with the mass of the halo. In~their case, the DM 
column density, $S$, was given by
\begin{equation}
\log S= 0.21 \log \frac{M_{halo}}{10^{10} M_{\odot}}+1.79,
\end{equation}
with S in $M_{\odot}/pc^2$. 
Cardone and Tortora \cite{CardoneTortora2010} 
showed that 
the Newtonian acceleration and~the column density correlate with different quantities: the visual luminosity $L_V$, the~effective radius $R_{eff}$, the~stellar mass $M_{\ast}$, and~the halo mass $M_{200}$ in agreement with~\cite{Boyarsky}, and~in disagreement with the D09 and~G09~results. 

Napolitano et al. \cite{Napolitano2010} found the non-existence of a universal surface density in early-type galaxies,  
while~\cite{2013MNRAS4291080D}, in~agreement with~\cite{Napolitano2010,CardoneTortora2010,Boyarsky}, found a correlation of the surface density with $M_{200}$, and~\cite{CardoneDelPopolo2012} found a a correlation between the Newtonian acceleration and the virial mass $M_{\rm vir}$. Similarly, from~the analysis of Saburova et al. 
\citep{Saburova2014} it is clear that the DM surface density correlates with several quantities, in~agreement with Zhou et al. \citep{Zhou2020} who reobtained the surface density with a Burkert profile and inferring the parameters through the Markov Chain Monte Carlo (MCMC) method.

In this paper, we extend the paper by Milgrom \citep{Milgrom2009} to spiral galaxies. 
As in \citep{Milgrom2009}, objects in the Newtonian regime are characterized by a constant surface density, while objects with lower acceleration does not show this behavior. In~summary, D09, and~G09 conclusion that the surface density, defined as the product of the parameters $\rho_0$, and~$r_0$ of the Burkert profile,  
is constant for every kind of galaxies, from~dwarfs to giants, is not in agreement with MOND result, apart than to all the results previously~mentioned.

The paper is organized as follows. In~Section~\ref{sec2}, we discuss the methods used to obtain the parameters of DM halos which were used in the current paper and in
~Section~\ref{sec3}, we discuss the results
. 

\section{Surface~Density}\label{sec2}
 
Mass and mass densities may be measured by means of their gravitational effects on the trajectories of massive particles. In~the following we follow \citep{Milgrom1986,Milgrom2009}.

If the acceleration field is given by $\bf g(r)$, this can be obtained from the potential $\phi$: $\bf g(r)=-\nabla$ $\phi$ which gives rise to the Poisson equation $\nabla^2 \phi = 4 \pi G \rho^*$. The~gravitational  mass density $\rho^* (\bf r)$ is given by
\begin{equation}
\rho^* (\mathbf{ r})=  \frac{1}{4 \pi G}\mathbf{ \nabla g(r)}.
\end{equation}   
In case the Newtonian dynamics are not applicable the quantity $\rho^* (\bf r)$ is not the true density $\rho(\bf r)$ giving rise to $\bf g(r)$. One then defines the phantom mass density given by $\rho_P (\mathbf{ r})= \rho^* (\mathbf{ r})- \rho(\bf r)$. 
The phantom density $\rho_P$ is considered as a real quantity but not observed to~date.

The quantity $\rho(\bf r)$ is given by
\begin{equation}
\rho(\mathbf{ r})= - \frac{1}{4 \pi G} \nabla [\mu ({\rm g}/a_0) \mathbf{ g}].
\label{eq:2}
\end{equation}
Ref.~\citep{Milgrom1986}, where $\mu(x)$ is the interpolating function, and~$a_0$ is the MOND acceleration constant.  
Equation~(\ref{eq:2}) does not fix $\bf g(r)$ uniquely for a given $\rho(\bf r)$. $\bf g$ is determined uniquely if in addition to Equation~(\ref{eq:2}) we require $\mathbf{g}=-\nabla \phi$. 
Equation~(\ref{eq:2}) can be written doing the derivatives
\begin{eqnarray}
\rho&=&\frac{-1}{4\pi G} \{\mathbf{ \nabla \cdot g} [\mu {\rm (g}/a_0)]+\mathbf{ g} \cdot \nabla [\mu {\rm (g}/a_0)]
\}= 
\nonumber\\
& & 
\rho^* \mu({\rm g}/a_0)-
\frac{1}{4 \pi G} {\mu'} 
({\rm g}/a_0) a_0^{-1} \bf g \cdot \nabla g.
\end{eqnarray} 
We then may write
\begin{eqnarray}
\rho_P (\bf r \rm) &=&\rho^*-\rho=\rho^*-\rho^* \mu +\frac{1}{4\pi G} \frac{\mu'}{a_0} \bf g \cdot \nabla g=
\nonumber \\
& &
\rho^*(1-\mu) +\frac{1}{4 \pi G} L \mu \rm \bf e_g \cdot \nabla g,
\end{eqnarray} 
where $L= \frac{\mu'}{\mu} {\rm g}/a_0$.
Recalling that $\rho^*=\rho+\rho_P$, we have
\begin{equation}
\rho_P=(\rho+\rho_P)(1-\mu) +\frac{1}{4 \pi G} L \mu \bf e_g \cdot \nabla g,
\end{equation}
and simplifying
\begin{equation}
\rho_P ({\bf r} )=\rho({\bf r}) (\frac{1}{\mu}-1)+\frac{1}{4 \pi G} L \rm \bf e_g \cdot \nabla g.
\end{equation}
This last equation can be written in terms of the potential recalling that ${\bf g}= - \nabla \phi$:
\begin{equation}
\rho_P ({\bf r \rm})=-\frac{1}{4 \pi G a_0} \frac{\mu'}{\mu}  \nabla |\nabla \phi|  \nabla  \phi+\rho ({\bf r}) (1/\mu-1).
\end{equation}

Defining $\mathcal{V}(x)= \int L(x) dx$, with~$x={\rm g}/a_0$, $L=\frac{\mu'}{\mu} x$, and~defining a vector $\bf e$ in the direction of $ \nabla  \phi$, we can write
\begin{equation}
\rho_P=\rho_{P_1}+\rho_{P_2}=\frac{-a_0}{4 \pi G} {\bf e} \cdot \nabla \mathcal{V} ( |\nabla  \phi|/a_0)+(1/\mu-1) \rho,
\label{eq:compl}
\end{equation}
and
\begin{equation}
\rho_{P_1}=\frac{-a_0}{4 \pi G} {\bf e} \nabla  \mathcal{V} (|\nabla  \phi|/a_0),
\label{eq:p1}
\end{equation}
\begin{equation}
\rho_{P_2}=(1/\mu-1) \rho.
\label{eq:p2}
\end{equation} 
 
In the case of a point mass, the~central surface density given by Equation~(\ref{eq:p1}) can be obtained integrating $\rho_{P}$ as
\begin{eqnarray}
\Sigma(0)&=&\int^{+\infty}_{-\infty} (\rho_{P_1}+\rho_{P_2}) dz= 2 \int^{\infty}_0 (\rho_{P_1}+(1/\mu-1) \rho) dz=
\nonumber\\
& &
\Sigma_M [\mathcal{V(\infty)}-\mathcal{V}(0)]+0=
\nonumber\\
& &
\Sigma_M \int^\infty_0 L(x)dx=a \Sigma_M,
\end{eqnarray}
where $\Sigma_M=\frac{a_0}{2 \pi G}$. Please note that the integral of $\rho_{P_2}$ is zero because inside the mass $\mu \simeq 1$. 
The value of $a$ depends on the interpolation function. For example, for $n=2$, $\mu=\frac{x}{(1+x^n)^{1/n}}$, we have $\frac{\pi}{2} \Sigma_M $ and for larger $n$ the value of $a$ tends to~1.

D09 gives a surface density integrating the Burkert profile\footnote{Please note that since the profile is spherically symmetric, integrating along an axis direction $z$ is equivalent to integrating twice along the positive radial direction}, given by
\begin{equation}
\Sigma_0= 2 \int \rho dr= 2 \int \frac{\rho_0 r_0^3}{(r+r_0) (r^2+r^2_0)}=\frac{\pi}{2} \rho_0 r_0=\frac{\pi}{2} \Sigma_c.
\end{equation}
If we call $\Sigma_c^*$ the MOND analog of $\Sigma_c$ we have, using the MOND to non-MOND surface density ratio $\lambda$ introduced in~\cite{Milgrom2009},
\begin{equation}
\Sigma_c^*=\frac{2\lambda}{\pi } \Sigma_M,
\end{equation}
with $\Sigma_M=138 \frac{a_0}{1.2 \times 10^{-8} \rm cm s^{-2}} M_\odot/pc^2$.

Our main goal is to extend the previous point mass calculation to the case of 
disk systems. As~we saw, Equation~(\ref{eq:p1}) may be applied to spherical systems, but it can also be applied 
to disk systems.
%
 
To calculate the integral of Equation~(\ref{eq:compl}) we will distinguish two cases, the~one in which $x \geq 1$ (acceleration above the MOND universal constant), and~the other in which $x \leq 1$. At~this stage it is worth pointing out that in our theoretical study we focus on galaxies which are clearly in the $x\gg1$ or $x\ll1$ regimes. Since any qualitative behavior in the intermediate regime will smoothly connect the two, we are not interested in the quantitative details of such regime and will imply the clear regimes when referring to either $x \geq 1$ or $x \leq 1$ in what follows. Let us consider the second term  
(Equation~\ref{eq:p2}),  $\rho_{P_2}=(1/\mu-1) \rho$.

The dimensionless acceleration, obtained from the ratio of the Newtonian acceleration ($\mu(g/a_0) g$) and $a_0$, is given by $a_N=({\rm g}/a_0)\mu({\rm g}/a_0)$. If~we use $n=1$, and~$\mu=\frac{x}{(1+x^n)^{1/n}}=\frac{{\rm g}/a_0}{1+{\rm g}/a_0}$, 
multiplying for $\mu$ we have
\begin{equation}
\mu=\frac{\mu {\rm g}/a_0}{\mu {\rm g}/a_0+\mu}=\frac{a_N}{a_N+\mu},
\end{equation} 
solving with respect to $\mu$ we have
\begin{equation}
\mu=-1/2a_N +1/2 \sqrt{a_N^2+4a_N}.
\end{equation}
Here we used $n=1$ to have an algebraically  simplified form of the acceleration. Using $n=2$ complicates the algebra but the following results do not~change.

For a double exponential disk, with~cylindrical radius $R$ and altitude $z$,
\begin{equation}
\rho=\rho_0 e^{-R/h_r-|z|/h_z},
\end{equation} 
the surface density is
\begin{equation}
\Sigma=\int_0^{\infty} \rho dz, 
\end{equation}
and the mass
\begin{equation}
M=\int_0^{\infty} 2 \pi R \Sigma dR= 2 \pi \rho_0 h_z h_r^2,
\end{equation}
while the dimensionless MOND acceleration ratio, in~a similar way as in~\cite{Milgrom2009}, reads (with $\Sigma_b$, the~baryonic surface density, approximated by an integrated constant volume density $\Sigma_b\simeq \int \rho_0 dz=\rho_0 h_z$)
\begin{equation}
a{_N}\simeq\frac{GM}{a_0 R{^2}}= \frac{G2 \pi \rho{_0} h{_z} h{_r}{^2}}{a{_0} R{^2}} \simeq \frac{\Sigma{_b}}{\Sigma{_M}} \frac{h{_r}{^2}}{R{^2}}\Mov{.}
\end{equation}
\Mov{The acceleration here is approximated using the spherical expression in cylindrical symmetry. Such an approximation can be evaluated using, for~instance, (\cite{BinneyTremaine2008}, Equation~2.165) to compare with the spherical expression and obtain an error within some percent in the inner disk and within 10\% in the outer disk. The~}
MOND interpolating function becomes
\begin{equation}
\mu=-1/2\,{\frac {\Sigma_{b}\,{{\it h_r}}^{2}}{\Sigma_{M}\,{R}^{2}}}+1/2\,
\sqrt {{\frac {{\Sigma_{b}}^{2}{{\it h_r}}^{4}}{{\Sigma_{M}}^{2}{R}^{4}
}}+4\,{\frac {\Sigma_{b}\,{{\it h_r}}^{2}}{\Sigma_{M}\,{R}^{2}}}}.
\label{eq:muu}
\end{equation}
Please note that the MOND acceleration being defined as a radial acceleration following~\cite{Milgrom2009}, the~gravitational potential is approximated with the spherical expression. Since in the inner parts of the system, the error of the approximation with respect to a thin disk is of the order of some percent and some 10 percent further away where MOND is valid, and~we are evaluating qualitative behaviors, we only introduce the cylindrical disk geometry at the level of the mass distribution. Then
\begin{eqnarray}
\frac{1-\mu}{\mu}
 \int_{-\infty}^{+\infty} \rho dz & \simeq & \frac{2(1-\mu)}{\mu} \Sigma_b \exp^{-R/h_r}
\nonumber \\
& =&
\frac{2(1-\mu)}{\mu} a_N \Sigma_M \frac{R^2}{h_r^2} \exp^{-R/h_r}
\nonumber\\
& =&
\frac{2(1-\mu)}{\mu} x \mu \Sigma_M \frac{R^2}{h_r^2} \exp^{-R/h_r}
\nonumber\\
&= &
2(1-\mu) x \Sigma_M \frac{R^2}{h_r^2} \exp^{-R/h_r}.
\end{eqnarray}
The full integral can be written as
\begin{equation}
F=F_1+F_2=\frac{2}{\pi}\Sigma_M \int_0^x L(x) dx + \frac{2}{\pi} 2(1-\mu) x \Sigma_M \frac{R^2}{h_r^2} \exp^{-R/h_r}.
\label{eq:tot}
\end{equation}
~\\
CASE $x \leq 1$
~\\

We recall that ${\rm g}=a_0 x$, and~in the case $x \leq 1$,  $\mu=x$, $a_N=({\rm g}/a_0) \mu=x^2$, and~from 
$a_N=x^2=\frac{\Sigma_b}{\Sigma_M} \frac{h_r^2}{R^2}$, $x=\sqrt{\frac{\Sigma_b}{\Sigma_M}}\frac{h_r}{R}$.

Concerning the first term $F_1$,
\begin{equation}
F_1=\frac{2}{\pi}\Sigma_M \int_0^x L(x) dx,
\end{equation} 
using $\mu=x/\sqrt{1+x^2}$, recalling that $L=\frac{\mu'}{\mu} x$, and~
$x=\sqrt{\frac{\Sigma_b}{\Sigma_M}}\frac{h_r}{R}$, one can obtain $F_1$ in terms of 
$\frac{\Sigma_b}{\Sigma_M}$, and~for small $x$ or 
small $\frac{\Sigma_b}{\Sigma_M}$
we obtain
\begin{eqnarray}
F_1&=&\frac{2}{\pi}\Sigma_M \int_0^x L(x) dx=\frac{2}{\pi}\Sigma_M \arctan{x}
\nonumber \\
& =&
\frac{2}{\pi} \Sigma_M \arctan{(\sqrt{\frac{\Sigma_b}{\Sigma_M}}\frac{h_r}{R})}.
\end{eqnarray}

Substituting Equation~(\ref{eq:muu}) in the right term $F_2$ of Equation~(\ref{eq:tot}) we obtain
\begin{eqnarray}
F_2&=& \frac{2}{\pi} 2
\frac {R\Sigma_{M}}{{\it h_r}} \left( 1+1/2\,{\frac {\Sigma_{b}\,{{
\it h_r}}^{2}}{\Sigma_{M}\,{R}^{2}}}-1/2\,\sqrt {{\frac {{\Sigma_{b}}^{
2}{{\it h_r}}^{4}}{{\Sigma_{M}}^{2}{R}^{4}}}+4\,{\frac {\Sigma_{b}\,{{
\it h_r}}^{2}}{\Sigma_{M}\,{R}^{2}}}} \right) 
\nonumber\\
& &\times
\sqrt {{\frac {\Sigma_{b}}
{\Sigma_{M}}}}
{{\rm e}^{-{\frac {R}{{\it h_r}}}}},
\end{eqnarray}
that at first order can be written as
\begin{equation}
F_2=(4/\pi) \,{\frac {{\it h_r}\,\Sigma_{M}}{R}{{\rm e}^{-{\frac {R}{{\it h_r}}}}}
\sqrt {{\frac {\Sigma_{b}}{\Sigma_{M}}}}}.
\end{equation} 

Then we have that
\begin{eqnarray}
F&=& \frac{2}{\pi} \Sigma_M \arctan{(\sqrt{\frac{\Sigma_b}{\Sigma_M}}\frac{h_r}{R})}+ 
\nonumber\\
& &
\frac{2}{\pi} 2
\frac {R\Sigma_{M}}{{\it h_r}} \left( 1+1/2\,{\frac {\Sigma_{b}\,{{
\it h_r}}^{2}}{\Sigma_{M}\,{R}^{2}}}-1/2\,\sqrt {{\frac {{\Sigma_{b}}^{
2}{{\it h_r}}^{4}}{{\Sigma_{M}}^{2}{R}^{4}}}+4\,{\frac {\Sigma_{b}\,{{
\it h_r}}^{2}}{\Sigma_{M}\,{R}^{2}}}} \right) 
\nonumber\\
& &\times
\sqrt {{\frac {\Sigma_{b}}
{\Sigma_{M}}}}
{{\rm e}^{-{\frac {R}{{\it h_r}}}}}.
\end{eqnarray}

At small $x$ $F_1$ tend to zero, and~$F$ is dominated by the second~term.

~\\
CASE $x \geq 1$
~\\

Concerning the case $x \geq 1$, we have again Equation~(\ref{eq:tot})
\begin{equation}
F=\frac{2}{\pi}\Sigma_M \int_0^x L(x) dx + (4/\pi) (1-\mu) x \Sigma_M \frac{R^2}{h_r^2} \exp^{-R/h_r}.
\end{equation} 
Recalling that $L=\frac{\mu'}{\mu} x$, $x=\frac{\Sigma_b}{\Sigma_M}\frac{h_r^2}{R^2}$
for $\mu=\frac{x}{\sqrt{1+x^2}}$
we have
\begin{eqnarray}
F_1&=&\frac{2}{\pi}\Sigma_M \int_0^{x} L(x) dx=\frac{2}{\pi} \Sigma_M \arctan{x}
\nonumber \\
&= & 
\frac{2}{\pi} \Sigma_M \arctan{(\frac{\Sigma_b}{\Sigma_M}\frac{h_r^2}{R^2})},
\end{eqnarray} 
and
\begin{eqnarray}
F_2&=&\frac{4}{\pi}\, \left( 1+1/2\,{\frac {\Sigma_{b}\,{{\it h_r}}^{2}}{\Sigma_{M}\,{R}^
{2}}}-1/2\,\sqrt {{\frac {{\Sigma_{b}}^{2}{{\it h_r}}^{4}}{{\Sigma_{M}}
^{2}{R}^{4}}}+4\,{\frac {\Sigma_{b}\,{{\it h_r}}^{2}}{\Sigma_{M}\,{R}^{
2}}}} \right) 
\nonumber\\
& &\times
x\Sigma_{M}\, {{\rm e}^{-{\frac {R}{{\it h_r}}}}} R^2/h_r^2,
\label{eq:o}
\end{eqnarray} 
so for $x \geq 1$, or~$\Sigma_b \geq \Sigma_M$, $x=\frac{\Sigma_b}{\Sigma_M}\frac{h_r^2}{R^2}$, 
we~have

%
%
\begin{eqnarray}
F &=& \frac{2}{\pi}\Sigma_M \arctan{(\frac{\Sigma_b}{\Sigma_M}\frac{h_r^2}{R^2})}+
\nonumber \\
& &
\frac{4}{\pi}\, \left( 1+1/2\,{\frac {\Sigma_{b}\,{{\it h_r}}^{2}}{\Sigma_{M}\,{R}^
{2}}}-1/2\,\sqrt {{\frac {{\Sigma_{b}}^{2}{{\it h_r}}^{4}}{{\Sigma_{M}}
^{2}{R}^{4}}}+4\,
{\frac {\Sigma_{b}\,{{\it h_r}}^{2}}{\Sigma_{M}\,{R}^{
2}}}} \right)
\nonumber\\
& &\times
\Sigma_{M}\,
{{\rm e}^{-{\frac {R}{{\it h_r}}}}} R^2/h_r^2.
\label{eq:finall}
\end{eqnarray}

For large $x$, $F_2$ tends to 0, O(1), and~$F_1 \simeq \Sigma_M$.

\begin{figure}[t]
\centering 
\includegraphics[width=15cm,angle=0]{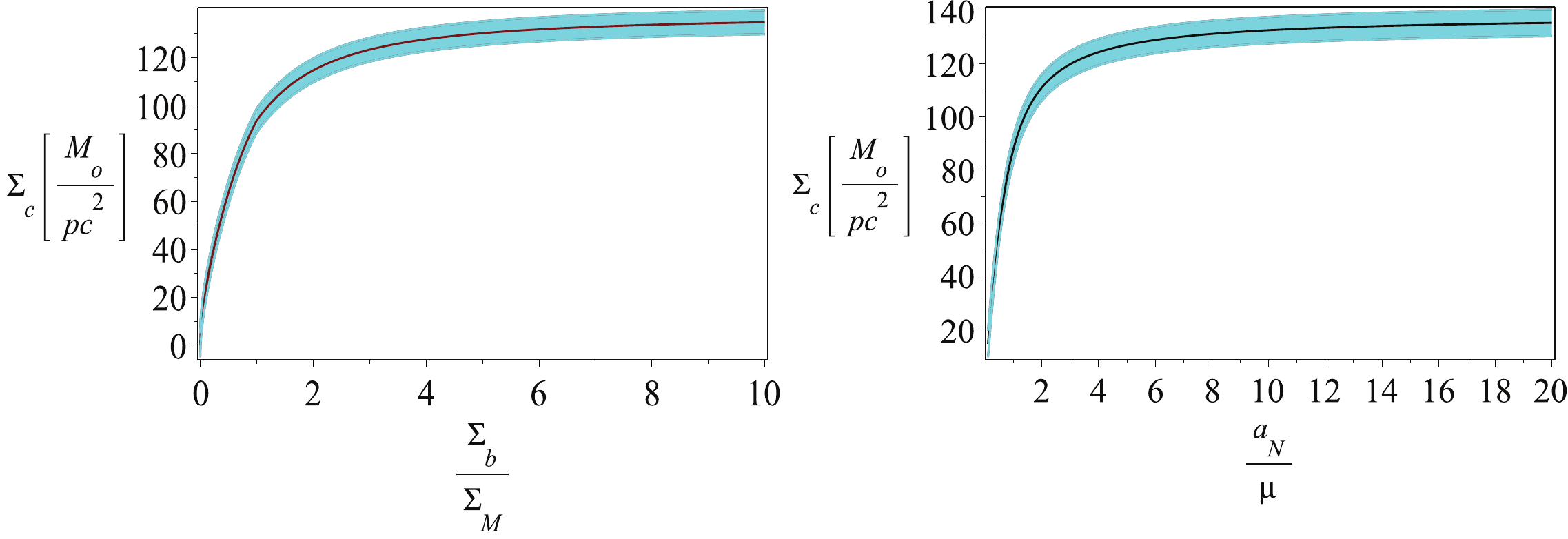}
 \caption[justified]{Left panel: the MOND column density in terms of the ratio $\Sigma_b/\Sigma_M$. Right panel: The MOND column density in terms of the ratio of the acceleration, $a_N$, and~$\mu$\Mov{. In~both panels, the~shaded regions bounded by dashed curves represent the $1\sigma$ evaluation induced by the approximations described in the text.}
}
 \label{fig:comparison}
\end{figure}

Assuming $h_r=R$, which corresponds to focusing at the edge of the central region of the galaxy where measurement of the central surface density is clearer, we plot in the left panel of Figure~\ref{fig:comparison} the $\Sigma_c^*$ in terms of $\Sigma_b/\Sigma_M$\Mov{, representing the $1\sigma$ errors of the evaluation, propagated from the spherical potential approximation, with~shaded areas bounded by dashed curves}.

The figure shows that there is a double trend of the surface density. At~small $\Sigma_b/\Sigma_M$, for~$R=h_r$, the~surface density increases as $\sqrt{\frac{\Sigma_b}{\Sigma_M}} \Sigma_M$. Going to larger $\Sigma_b/\Sigma_M$, the~plot flattens till when, at~large $\Sigma_b/\Sigma_M$ tends to $\Sigma_M$.   \Mov{The previous results, plotted in the left panel of Figure~\ref{fig:comparison}, rely on assuming that the behavior at intermediate values of $x$ and $\Sigma_b/\Sigma_M$ should remain smooth. This is vindicated by the exact calculation below, leading to Equation~\eqref{eq:integral} and the plotting of Figure~\ref{fig:comparison}'s right panel, and~which is independent of the previous approximations.}

Apart the dependence of $\Sigma_c^*$ on the ratio $\Sigma_b/\Sigma_M$, the~dependence from the acceleration $a_N/\mu$ can be obtained by the relation
\begin{eqnarray}
F &=& \frac{2}{\pi}\Sigma_M \int_0^x L(x) dx + \frac{4}{\pi}(1-\mu) x \Sigma_M \frac{R^2}{h_r^2} \exp^{-R/h_r}
\nonumber \\
& =&
\frac{2}{\pi} \Sigma_M \arctan{x} +\frac{4}{\pi}(1-\mu) x \Sigma_M \frac{R^2}{h_r^2} \exp^{-R/h_r},\label{eq:integral}
\end{eqnarray}
with the previous definition of $\mu=\frac{x}{\sqrt{1+x^2}}$, and~$L(x)=x \frac{\mu'}{\mu}$, and~assuming $h_r=R$, we plot in the right panel of Figure~\ref{fig:comparison} the relation $\Sigma_c^*$ versus $a_N/\mu$. 
 
As in the left panel of Figure~\ref{fig:comparison}, the~plot shows again a double trend of the surface density. At~small $x={\rm g}/a_0$, for~$R=h_r$, the~surface density increases as $ x\Sigma_M$. Going to larger $x$, the~plot flattens and reach the asymptote $\Sigma_M$.  \Mov{The behavior in the right panel of Figure~\ref{fig:comparison} deriving from Equation~\eqref{eq:integral}, it is exact. As~the left panel displays a similar asymptotic behavior to the right panel for both $a_N\ll a_0$ and $a_N\gg a_0$, and~we expect a smooth junction between the two regions, the~plotting of the smooth junction is justified.}

The previous plots and the result of the present paper can be summarized as follows, and~represent the main result of the paper. 
The previous results show that the claim of existence of a quasi-universal surface density in agreement with D09, and~G09, is in disagreement with MOND, which predicts that for small values of baryonic surface density $\Sigma_b$, or~small acceleration, the~surface density decreases below the quasi-universal value. This result is also in agreement with~\cite{Napolitano2010,Boyarsky,2013MNRAS4291080D,CardoneTortora2010,CardoneDelPopolo2012,Saburova2014,Zhou2020}, namely the surface density is not universal. The~D09 and~G09 claims of the existence of a quasi-universal value of the surface density for all galaxies (considering low surface density ones) is denied also from the existence of galaxies having values much smaller than $\Sigma_M$. As an example, NGC 3741 according to \citep{Begum2005} has a value $56 M_{\odot}/pc^2$ much smaller than the quasi-universal value, or~similarly KK98 250, KK98 251 has shown by \citep{Begum2004} have values $56 M_{\odot}/pc^2$, and~$66 M_{\odot}/pc^2$. In~a future paper, we want to use the SPARC sample to determine the values of the surface density and compare with our model. 
One question that could arise, is why does D09 claim the universality of the surface density? Probably this is due to the fact they had few galaxies at small surface density, and~they were plagued by errors. In~their Figure 2, 
from magnitude $-$15 to 7, they had just seven galaxies. One is NGC 2137, whose surface density errors were probably overestimated. The~others are  
dwarf spheroidals, that as reported by the same D09 is beset with uncertainties in the model assumptions, leading to non-unique results. Concerning high surface density systems, the~D09 result is in agreement with MOND, namely there is a quasi-universal surface~density. 

The MOND prediction of the non-existence of a universal surface density on a large range of surface densities or acceleration, is in agreement with several studies~\cite{Napolitano2010,Boyarsky,2013MNRAS4291080D,CardoneTortora2010,CardoneDelPopolo2012,Saburova2014,Zhou2020}.   
which disagrees with D09~result. 
 
\section{Discussion}  \label{sec3}

We extended a paper of Milgrom related to the prediction of the existence of a quasi-universal 
central density surface density of dark halos. According to this study, systems characterized by a mean acceleration larger than MOND typical acceleration $a_0$, namely if they are in Newtonian regime, are characterized by a quasi-universal surface density, whose value is proportional to $\Sigma_M=\rm \frac{a_0}{2 \pi G}=138 \frac{a_0}{1.2\times 10^{-8} cm s^{-2}} M_\odot/pc^{-2}$. This claim is in agreement with those of D09 and~G09. Milgrom~\citep{Milgrom2009} 
also calculated in the case of general (non-disky
) systems with constant density, an~approximated value for the surface density, showing that for systems of low surface density there is no longer a quasi-universal value, but~the studied systems are characterized by lower values of the surface density, which contradicts D09. In~the present paper, the~calculation of Milgrom, set from a point mass model, was extended to spiral galaxies, modeled with a cylindrically symmetric double exponential disk. We found a quasi-universal surface density for Newtonian systems, and~confirmed smaller values of the surface density for low acceleration systems in our extended cylindrical~configuration.

\vspace{6pt} 



\section*{Author contributions}
All authors have contributed equally in this work. All authors have read and
	agreed to the published version of the manuscript.

\section*{Funding}
{MLeD acknowledges financial support by the Lanzhou University starting fund, the
	Fundamental Research Funds for the Central Universities (Grant No. lzujbky-2019-25), the National
	Science Foundation of China (grant No. 12047501), and the 111 Project under Grant No. B20063.}

\section*{Institutional review}
{N/A.}

\section*{Informed consent}
{N/A.}

\section*{Data availability}
{N/A.}

\section*{Acknowledgments}
{The authors wish to thank Francesco Pace for some calculations.}

\section*{Conflicts of interest}
{The authors declare no conflict of interest.} 

\end{document}